# Iron Displacements and Magnetoelastic Coupling in the Spin-Ladder Compound BaFe$_2$Se$_3$

J. M. Caron[1], J. R. Neilson[1], D. C. Miller[1], A. Llobet[2], and T. M. McQueen[1,†]

[1]Institute for Quantum Matter, Department of Chemistry, and Department of Physics and Astronomy, The Johns Hopkins University, Baltimore, MD 21030
[2]Los Alamos National Laboratory, Lujan Neutron Scattering Center, MS H805, Los Alamos, NM 87545

We report long-range ordered antiferromagnetism concomitant with local iron displacements in the spin-ladder compound BaFe$_2$Se$_3$. Short-range magnetic correlations, present at room temperature, develop into long-range antiferromagnetic order below $T_N$ = 256 K, with no superconductivity down to 1.8 K. Built of ferromagnetic Fe$_4$ plaquettes, the magnetic ground state correlates with local displacements of the Fe atoms. These iron displacements imply significant magnetoelastic coupling in Fe$X_4$-based materials, an ingredient hypothesized to be important in the emergence of superconductivity. This result also suggests that knowledge of these local displacements is essential for properly understanding the electronic structure of these systems. As with the copper oxide superconductors two decades ago, our results highlight the importance of reduced dimensionality spin ladder compounds in the study of the coupling of spin, charge, and atom positions in superconducting materials.

The propensity of iron to form magnetically ordered ground states made the 2005 discovery [1-9] of high temperature superconductivity in iron-based materials very surprising. Built of edge-sharing Fe$X_4$ ($X$ = anion) tetrahedra, these materials share many structural characteristics with the copper oxide superconductors, including a layered, two-dimensional structure and a nearly square arrangement of metal atoms. These similarities led to speculation that magnetic correlations, important in the copper oxides, play a key role in the appearance of superconductivity in the high-$T_c$ iron compounds. Even with many recent experimental and theoretical studies, the mechanism of superconductivity in iron-based materials remains elusive. Uncertainties persist over the nature of the superconducting gap, the relationship between the iron pnictides and iron chalcogenides, and the relevance of magnetic, structural, and charge degrees of freedom on the resulting superconductivity [10-25].

Here, we present a systematic study of the structural and magnetic properties of the spin-ladder iron chalcogenide, BaFe$_2$Se$_3$, using neutron powder diffraction (NPD), neutron pair-distribution function (n-PDF) analysis, and magnetization measurements. Consistent with the previous literature, we find that the structure consists of double chains of edge-sharing FeSe$_4$ tetrahedra separated by Ba atoms (Fig. 1) [26]. These double chains are cut out of the two dimensional layers found in the superconducting iron compounds, and are akin to the two-legged spin-ladders in copper oxides such as SrCu$_2$O$_3$ [27].

BaFe$_2$Se$_3$ exhibits a number of remarkable properties derived from these spin-ladders. Short-range magnetic correlations ($\xi$ ~ 35 Å) are observed from diffuse scattering in NPD at room temperature, above $T_N$. Surprising for such a quasi-1D system in which there are no covalent linkages between spin-ladders, long-range antiferromagnetic order develops below $T_N$ = 256 K. The uniquely determined magnetic structure is built of ferromagnetic Fe$_4$ plaquettes; spins align perpendicular to the plane of the ladders, with an alternating spin direction between adjacent plaquettes.

Neutron pair distribution function analysis unequivocally shows displacements of the Fe atoms from their ideal positions within each ladder, and these distortions appear to increase when antiferromagnetic order develops, indicative of significant magnetoelastic coupling. Furthermore, as in other materials in which local distortions are present (e.g. charge density wave systems), these atom displacements likely have a significant impact on the electronic structure by modulating the local wavefunctions. Taken together, these data and analysis intimate a relationship between positional movements of the Fe atoms and the resulting electronic and magnetic properties of Fe$X_4$-based materials [28, 29].

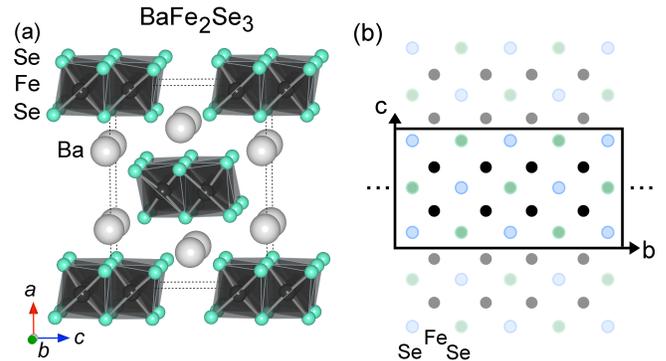

**Figure 1**. (a) The structure of BaFe$_2$Se$_3$ consists of double chains of edge-sharing FeSe$_4$ tetrahedra, which extend in and out of the page. The chains are well-separated from each other by Ba ions. (b) The double chains in BaFe$_2$Se$_3$ are cut outs of the two dimensional layers of edge-sharing Fe$X_4$ tetrahedra found in the iron-based superconductors. Each double chain consists of pairs of Fe atoms ('rungs') tiled in a chain along one direction, forming a ladder structure.

All measurements were performed on powder samples synthesized from the elements in a multi-step process. Vacuum re-melted and polished iron pieces were reacted with selenium in a 2:3 molar ratio in a double sealed quartz tube at 750 °C [28] until the Se vapor disappeared, and then quenched. The resulting multiphase Fe-Se mixture was mixed in a stoichiometric ratio with redistilled barium pieces and heated as pressed pellets in alumina crucibles in evacuated quartz ampoules at successively higher temperatures between 500 and 750 °C, with furnace cooling and intermediate regrinding between each overnight heating, until the product was single phase by laboratory x-ray diffraction. NPD data were collected on powder samples loaded into vanadium cans at temperatures between 5 K and

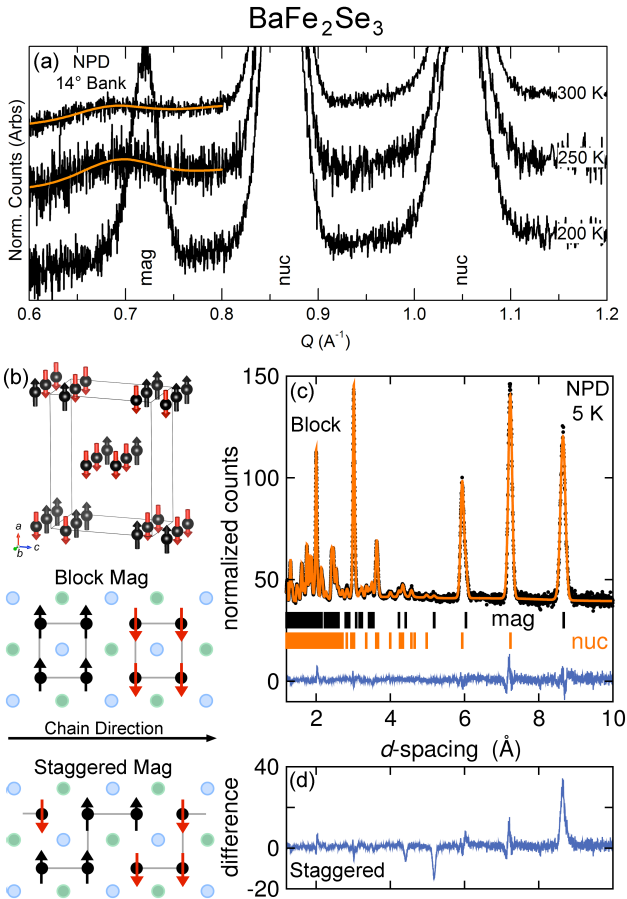

**Figure 2**. (a) In addition to sharp nuclear Bragg reflections, NPD data at 300 K exhibit a broad diffuse scattering peak attributable to magnetic correlations. In their stead, sharp magnetic Bragg peaks corresponding to long-range magnetic order develop by 200 K. (b,c) The magnetic structure that best fits the NPD data at 5 K is obtained with the all-block configuration with spins aligned along the crystallographic $a$ axis. (d) An all staggered configuration provides a significantly worse fit to the data, indicated by the systematic discrepancies in the difference between the data and the predicted intensities.

300 K on the time-of-flight instruments HIPD and NPDF at Los Alamos Neutron Science Center (LANSCE), Los Alamos National Laboratory (LANL). Magnetic structure determination was performed by Rietveld refinement of the NPD data using FullProf [30] combined with representational analysis from SARA$h$ [31], following initial refinements using GSAS and EXPGUI [33]. A small, <1% iron impurity phase was included in the final refinements.
Pair-distribution function analysis was performed on the NPD data and $G(r)$ were extracted with $Q_{max}$ = 35 Å$^{-1}$ using PDFgetN [32]. Least-squares refinements to the pair-distribution functions were performed using PDFgui after refining the instrumental parameters $Q_{damp}$ = 0.0068 and $Q_{broad}$ = 0.0145 from data collected on polycrystalline Si. Rietveld analysis of the data collected on NPDF to the average nuclear structure was performed using GSAS [33]. DC magnetization measurements were performed between 1.8 K and 300 K using a Quantum Design Physical Properties Measurement System.

At 300 K, all sharp Bragg reflections in the NPD pattern are described by the previously reported nuclear structure [26]. There is an additional broad feature, centered at $Q \sim 0.7$ Å$^{-1}$, that is not described by the simple nuclear cell. This weak peak is likely magnetic in origin, as it disappears when long-range magnetic order develops, only occurs where the form factor for magnetic scattering is the highest, and is in close proximity to the long range elastic magnetic scattering peak. Approximate correlation lengths extracted from the peak widths are on the scale of $\xi \sim 35$ Å. Previous Mössbauer studies on BaFe$_2$Se$_3$ did not reveal any local magnetic order above 250 K [34], suggesting that the diffuse scattering is due to short range magnetic correlations (SRC) rather than static short-range magnetic order, and that the timescale of the fluctuations is slower than the neutron timescale (~10$^{-13}$ s) but faster than the Mössbauer one (~10$^{-7}$ s). However, more detailed experiments are necessary to provide a complete understanding of the SRC at 300 K in BaFe$_2$Se$_3$.

Long-range magnetic order (LRO) replaces SRC as the system is cooled below $T_N$ = 256 K. The diffuse scattering at $Q \sim 0.7$ Å from SRC disappears and is replaced by a sharp magnetic Bragg peak at $Q = 0.728$ Å from LRO. At the same time, additional magnetic Bragg peaks appear, and the magnetic LRO state is mostly developed by 200 K. The nature of the LRO magnetic state was determined conclusively from NPD data using representational analysis combined with magnetic Rietveld refinements. The longest propagation vector capable of describing the magnetic structure is $\vec{k} = \langle ½, ½, ½ \rangle$, determined by indexing Bragg reflections with magnetic contributions, which were identified by intensity differences between 200 K and 300 K. A full representational analysis to determine possible irreducible representations (irreps) and basis vectors (BVs) to describe the magnetic structure was carried out using SARA$h$. There are a total of 24 possible basis vectors evenly split between two irreps, $\Gamma_1$ and $\Gamma_2$. Each irrep describes four of the eight magnetically distinct atoms within the unit cells, and both irreps join as a corepresentation in order to describe all eight unique magnetic atoms. They are labeled below following the scheme of SARA*h* and Kovalev and detailed in Table I [35]. Within each irrep, the 12 BVs describe the relationship of two atom pairs within distinct chains, with moments pointing along the *a*, *b*, and *c* axes.

**Table I.** Real components along the unit cell axes of the non-zero basis vectors used to describe the block magnetic structure in *Pnma* with $\vec{k} = \langle ½, ½, ½ \rangle$ and their approximate positional coordinates labeled using the convention of SARA$h$ [31] and Kovalev [35]. The magnetic structure, $\Gamma_{mag}$ is defined with one Fourier coefficient ($c$) for all four basis vectors, $\Gamma_{mag} = \sum_{4,10,16,22} c\psi_i$.

| BVs $\Psi_i$ | Atom | Basis vector components | | | Positional coordinates | | |
|---|---|---|---|---|---|---|---|
| | | $m_a$ | $m_b$ | $m_c$ | x | y | z |
| $\Psi_4$ | Fe4 | 1 | 0 | 0 | $\bar{x}$ + ½ | $\bar{y}$ | z |
| | Fe6 | -1 | 0 | 0 | $\bar{x}$ + ½ | y + ½ | z + ½ |
| $\Psi_{10}$ | Fe1 | 1 | 0 | 0 | x | y | z |
| | Fe7 | -1 | 0 | 0 | x | $\bar{y}$ + ½ | z |
| $\Psi_{16}$ | Fe3 | -1 | 0 | 0 | $\bar{x}$ | y + ½ | $\bar{z}$ |
| | Fe5 | -1 | 0 | 0 | $\bar{x}$ | $\bar{y}$ | $\bar{z}$ |
| $\Psi_{22}$ | Fe2 | -1 | 0 | 0 | x + ½ | $\bar{y}$ + ½ | $\bar{z}$ + ½ |
| | Fe8 | -1 | 0 | 0 | x + ½ | y | $\bar{z}$ + ½ |

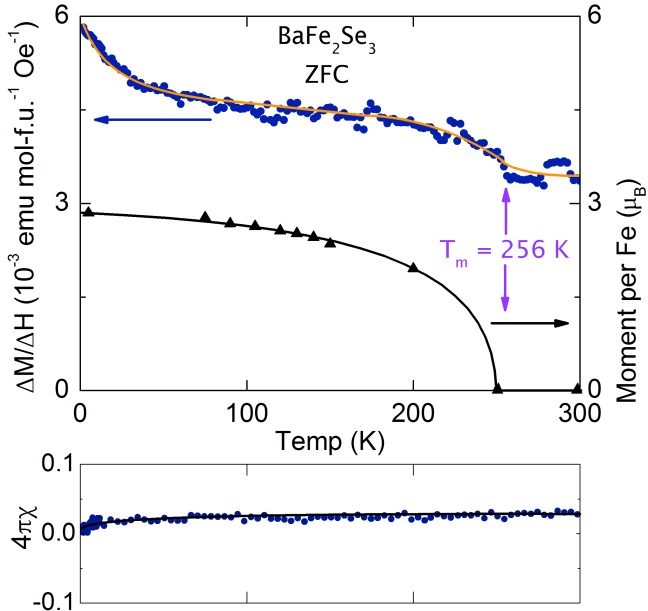

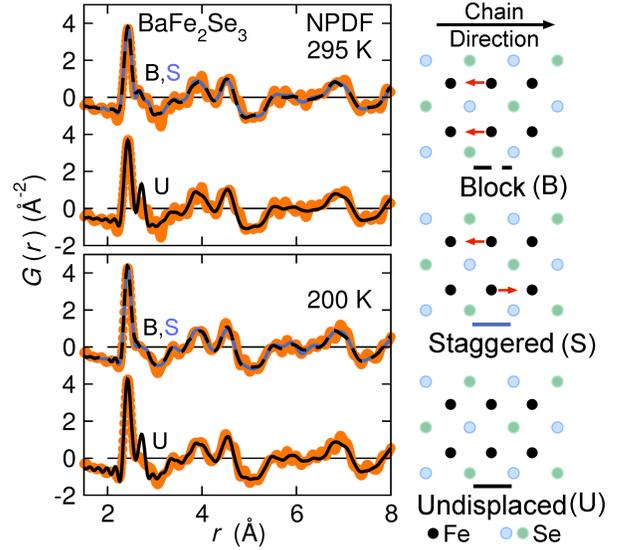

**Figure 3**. (a) The temperature dependence of the magnetic susceptibility for BaFe$_2$Se$_3$ shows a discontinuity at 256 K, near the same temperature that long-range magnetic order develops (indicated by the moment per Fe, extracted from Rietveld refinements of NPD data). Error bars for the NPD moments are less than the size of the symbols, and the scatter in the ΔM/ΔH data is from the subtraction process used to remove the ferromagnetic contribution from <1% Fe metal. (b) A low field ($\mu_0 H = 10$ Oe), zero field cooled magnetization curve for BaFe$_2$Se$_3$ conclusively demonstrates that the material is not a bulk superconductor above 1.8 K (perfect diamagnetism corresponds to -1 on the $y$ axis). The lines are guides to the eye.

**Table II.** Average crystallographic structure parameters for BaFe$_2$Se$_3$ at 5 K from neutron diffraction data recorded on the time-of-flight instrument HIPD. Space group *Pnma*. $a = 11.8834$ Å, $b = 5.4141$ Å, $c = 9.1409$ Å. A ~1% Fe impurity was included. All crystallographic sites are fully occupied.

| Atom | Wyckoff Position | $U_{iso}$ / Å$^2$ |
|---|---|---|
| Ba1 | 4c (0.186(3), ¼, 0.518(7)) | 0.012(14) |
| Fe1 | 8d (0.493(2), 0.002(4), 0.353(2)) | 0.005(6) |
| Se1 | 4c (0.355(2), ¼, 0.233(3)) | 0.003(6) |
| Se2 | 4c (0.630(2), ¼, 0.491(4)) | 0.003(6) |
| Se3 | 4c (0.402(2), ¼, 0.818(3)) | 0.003(6) |

For a given moment direction (or mixture of directions), and assuming each Fe atom carries the same moment parallel or antiparallel to a common spin axis, the four sets of BVs describe sixteen magnetic arrangements within a spin ladder. These sixteen are symmetry equivalents of three basic magnetic patterns: a block configuration consisting of ferromagnetic Fe$_4$ plaquettes tiled antiferromagnetically along the ladder, a staggered configuration with ferromagnetic diagonal double stripes that are also tiled antiferromagnetically (Fig. 2(b)), and a weave configuration that is analogous to the block configuration but with two spins up and two spins down per block (not shown).

Placing one of these sixteen units on each of the two independent spin ladders per unit cell gives 256 possible arrangements. These describe 64 unique magnetic structures, with the remainder being related by choice of origin. Each of these 64 was tested by Rietveld refinement against the NPD data, each with an arbitrary spin axis. The best fit to the data, with $R_{mag} = 4.35\%$, (HIPD 40° bank; 1.15-11 Å) contains spin ladders with the block magnetic structure illustrated by spins aligned with the crystallographic *a* axis, parallel or antiparallel to the double chains (Fig. 2(c); $\Gamma_{mag} = \sum_{4,10,16,22} c \psi_i$). The next best magnetic refinement was obtained with a mixture of block and staggered configurations, with $R_{mag} = 17.3\%$ ($\Gamma_{mag} = \sum_{7,10,16,22} c \psi_i$). The best staggered arrangement, proposed as a possible magnetic structure in Ref. [36], poorly describes our data, with a large $R_{mag} = 42.1\%$ and visually present errors in the residual (Fig. 2(d); $\Gamma_{mag} = \sum_{1,7,16,22} c \psi_i$). Thus we unambiguously identify the magnetic structure of BaFe$_2$Se$_3$ as ferromagnetic Fe$_4$ plaquettes tiled antiferromagnetically along each ladder. Crystallographic parameters for the final Rietveld refinement of the NPD data at 5 K are given in Table II.

**Figure 4**. Local structure analysis with NPDF unambiguously illustrates that the Fe atoms within the spin-ladder are each displaced by ~ 0.05 Å from their ideal positions, in either a block or staggered configuration. These local iron displacements, not visible in the average crystal structure, increase in magnitude as magnetic order develops, indicating strong magnetoelastic coupling in the iron chalcogenide family of materials.

The temperature evolution of the magnetic moment per iron is shown in Fig. 3(a). The moment extracted from the NPD refinements shows a sharp upturn below 250 K and saturates near 2.80(8) $\mu_B$/Fe by 5 K. This is smaller than the 4 $\mu_B$/Fe$^{2+}$ expected for high spin iron in tetrahedral coordination but larger than the 2 $\mu_B$/Fe$^{2+}$ for the low-spin case. The reduction in moment observed may be due to local metal-metal bonding, and/or due to proximity to a local-to-itinerant electronic transition. Similar low moments are common in related iron-based materials [12,16].

The magnetic susceptibility, calculated assuming $\chi \approx \Delta M/\Delta H$ from the difference between magnetization data collected at $\mu_0H$ = 1 T and 2 T to eliminate a ferromagnetic contribution from <1% impurity iron, also exhibits a discontinuity indicative of a magnetic transition at $T_N$ = 256 K. By these data as well, BaFe$_2$Se$_3$ undergoes a transition from SRC to a LRO state at this temperature. The upturn at low $T$ in the susceptibility, not present in the moment extracted from NPD refinements, is fit by the Curie-Weiss law for paramagnetic spins, and accounts for <5% impurity spins in the material, likely arising from defects within BaFe$_2$Se$_3$. Superconductivity in BaFe$_2$Se$_3$, suggested in a previous report [36], could also result in such an upturn in $\Delta M/\Delta H$ due to greater diamagnetic shielding at the lower applied field. However, a sensitive zero field cooled dc test for superconductivity under an applied field of H = 10 Oe, shown in Fig. 3(b), unambiguously demonstrates that BaFe$_2$Se$_3$ is not superconducting above 1.8 K.

Commensurate with the block magnetic order, we find displacements of the Fe atoms from their ideal position within each spin ladder by n-PDF analysis, reminiscent of a spin-Peierls distortion. While resulting in an excellent Rietveld fit to the NPD data, a structure with the Fe atoms forming an 'ideal' ladder (*i.e. y* = 0 for Fe in *Pnma*, and all Fe-Fe distances along the ladder equivalent) is a poor fit to the local structure from n-PDF analysis at 295 K (Fig. 4), with $R_w$ = 36.8%. An adequate fit to the local structure is only obtained when the Fe atoms are displaced by ~0.05 Å along the chain direction from the ideal position. The reported average crystallographic symmetry, *Pnma*, allows this to happen by staggering the Fe atoms. Such a staggering is indistinguishable from the undisplaced model in the NPD Rietveld refinements ($R_F$ = 5.34% versus $R_F$ = 5.44%; HIPD 153° bank; or $R_F$ = 7.73% versus $R_F$ = 7.72%; NPDF 148° bank), but considerably improves the fit to the n-PDF analysis (Displaced: $R_w$ = 27.4% *vs.* Undisplaced: $R_w$ = 36.8% %, Fig. 4). A second possibility is that the Fe atoms both displace in the same direction and form a crystallographic block structure, the spatially coherent average symmetry of which is $Pmc2_1$ (a subgroup of *Pnma* removing the *n* glide staggering the Fe atoms). However, in the local structure analysis, the *n* glide is removed from the Fe sites, and the iron atoms are constrained to displace in the same directions along the chain from their ideal position, while the Se atoms are allowed to relax according to *Pnma* (their positions along the chain are special); this block model is statistically indistinguishable from the staggered model in *Pnma* on the basis of either our n-PDF analysis ($R_w$ = 28.1%, Fig. 4) or Rietveld refinements of NPD data ($R_F$ = 4.48% versus $R_F$ = 4.22%, not significant given the number of extra free parameters by the Hamilton *R*-ratio test [37]).

Both distortion models (stagger or block) illustrate a temperature dependence of the Fe–Fe distances: the displacements increase upon cooling and are a direct consequence of magnetoelastic coupling. From least-squares refinement to the local n-PDF analysis, the two Fe-Fe distances along the chain at 295 K are 2.616(2) Å and 2.832(2) Å; the distances at 200 K are 2.593(1) Å and 2.840(1) Å. One Fe–Fe distance contracts while the other expands from 295 K to 200 K. Rietveld analysis of NPD data is not sensitive to these subtle displacements; the Fe–Fe lengths are independent of the cell parameters and only modulate subtleties of the Bragg peak intensities.

Given the block nature of the LRO magnetic structure, the block structure of Fe distortions seems more likely in the light of magnetoelastic coupling. Such coupling is reminiscent of the spatially coherent distortions giving rise to the well-known tetragonal-to-orthorhombic phase transition in the iron pnictides. In this case, the short-range magnetic correlations established above 295 K coexist with displacements of the iron positions. Whichever the direction or extent of spatial coherence of the distortions, the n-PDF analysis of the total scattering profile unambiguously shows local displacements of the Fe atoms; the magnitude of the distortion increases as magnetic long-range order develops. This unequivocally demonstrates significant magnetoelastic coupling in this spin-ladder iron chalcogenide.

Our results reveal the existence of local displacements in the Fe atoms above $T_N$ and their evolution during the development of magnetic order – i.e. magnetoelastic coupling – in iron-based materials, and call for further study of these structural-magnetic correlations. Such displacements, though small, have a significant impact on the electronic structure due to rearrangement of electrons near the Fermi level. Similar local offsets have been observed in FeSe and FeSe$_{1-x}$Te$_x$, with the presence or absence of superconductivity being very sensitive to their pattern: local distortions are important in the development of superconductivity [25,29,38]. As BaFe$_2$Se$_3$ does not superconduct, the magnetoelastic coupling alone is insufficient for electron condensation; we postulate that, like the spin-ladder copper oxide materials [39], this is due to reduced electronic bandwidth from the reduced dimensionality of a double chain versus complete layer. Our results highlight not only the importance of reduced dimensionality spin ladder compounds in the study of the coupling of spin, charge, and atom positions in superconducting materials, but also show that local metal atom displacements are ubiquitous in Fe$X_4$-based materials and must be included when trying to understanding the resulting magnetism and superconductivity.

TMM acknowledges useful discussions with O. Tchernyshyov and C. Broholm. Research supported by the U.S. Department of Energy, Office of Basic Energy Sciences, Division of Materials Sciences and Engineering under Award DE-FG02-08ER46544. This work has benefited from the use of HIPD and NPDF at the Lujan Center at Los Alamos Neutron Science Center, funded by DOE Office of Basic Energy Sciences. Los Alamos National Laboratory is operated by Los Alamos National Security LLC under DOE Contract DE-AC52-06NA25396. The upgrade of NPDF has been funded by NSF through grant DMR 00-76488.

† E-mail: mcqueen@jhu.edu
[1] Y. Kamihara, *et al. J. Am. Chem. Soc.* **128**, 10012-10013 (2006).
[2] Y. Kamihara, *et al. J. Am. Chem. Soc.* **130**, 3296-3297 (2008).
[3] F.C. Hsu, *et al. Proc. Nat. Acad. Sci.* **105**, 14262-14264 (2008).
[4] M. Rotter, M. Tegel, and D. Johrendt. *Phys. Rev. Lett.* **101**, 107006 (2008).

[5] S. Medvedev, *et al. Nat. Mat.* **8**, 630-633 (2009).
[6] K. Sasmal, *et al. Phys. Rev. Lett.* **101**, 107007 (2008).
[7] Y. Mizuguchi, *et al. App. Phys. Lett.* **93**, 152505 (2008).
[8] D.R. Parker, *et al. Chem. Comm.*, 2189-2191 (2009).
[9] J.H. Tapp, *et al. Phys. Rev. B* **78**, 060505 (2008).
[10] C.L. Chien, *et al. Nature* **453**, 1224-1227 (2008).
[11] Y.M. Qiu, *et al. Phys. Rev. Lett.* **103** (2009).
[12] A.D. Christianson, *et al. Nature* **456**, 930-932 (2008).
[13] V. Cvetkovic and Z. Tesanovic. *Europhys. Lett.* **85** (2009).
[14] H.Q. Yuan, *et al. Nature* **457**, 565-568 (2009).
[15] V. Cvetkovic and Z. Tesanovic. *Phys. Rev. B* **80** (2009).
[16] C. Cruz, *et al. Nature* **453**, 899-902 (2008).
[17] V.B. Zabolotnyy, *et al. Nature* **457**, 569-572 (2009).
[18] H.Q. Yuan, *et al. Nature* **457**, 565-568 (2009).
[19] X.H. Chen, *et al. Nature* **459**, 64-67 (2009).
[20] M. Greven, *et al. Nat. Phys.* **5**, 873-875 (2009).
[21] M.M. Qazilbash, *et al. Nat. Phys.* **5**, 647-650 (2009).
[22] I.I. Mazin. *Nature* **464**, 183-186 (2010).
[23] J.A. Kang and Z. Tesanovic. *Phys. Rev. B* **83** (2011).
[24] Z.Y. Yin, Z. P., K. Haule, and G. Kotliar. *Nat. Phys.* **7**, 294-297 (2011).
[25] T.M. McQueen, *et al. Phys. Rev. Lett.* **103**, 057002 (2009).
[26] H.Y. Hong and Steinfin, H. *J. Sol. St. Chem.* **5**, 93-& (1972).
[27] K. Ishida, *et al. J. Phys. Soc. Jpn.* **63**, 3222-3225 (1994).
[28] T.M. McQueen, *et al. Phys. Rev. B* **79**, 014522 (2009).
[29] D. Louca, *et al. Phys. Rev. B* **84**, 054522 (2011).
[30] J. Rodriguez-Carvajal. *Physica B* **192**, 55-69 (1993).
[31] A.S. Wills. *Physica B* **276**, 680-681 (2000).
[32] P.F. Peterson, *et al. J. Appl. Cryst.* **33**, 1192 (2000).
[33] A.C. Larson and R.B. Von Dreele, Los Alamos National Laboratory Report LAUR 86-748 (2000), and B. H. Toby. *J. Appl. Cryst.* **34**, 210-213 (2001).
[34] Reissner, M., Steiner, W., and Boller, H. *Hyperfine Interactions C: Proceedings* **5**, 197-200 (2002).
[35] O.V. Kovalev, H.T. Stokes, and D.M. Hatch. *Representations of the crystallographic space groups : irreducible representations, induced representations, and corepresentations*. 2nd. ed. 1993, Yverdon, Switzerland ; Langhorne, Pa.: Gordon and Breach. xiv, 390 p.
[36] A. Krzton-Maziopa, *et al.*, *J. Phys.: Cond. Mat.* **23**, 402201 (2011).
[37] W.C. Hamilton. *Acta Crystallographica* **18**, 502- (1965).
[38] D. Louca, *el al.*, *Phys. Rev. B* **81**, 134524 (2010).
[39] Z. Hiroi and M. Takano. *Nature* **377**, 41-43 (1995).